# Doping – dependent Irreversible Magnetic Properties of Ba(Fe$_{1-x}$Co$_x$)$_2$As$_2$ Single Crystals

R. Prozorov, M. A. Tanatar, E. C. Blomberg, P. Prommapan,

R. T. Gordon, N. Ni, S. L. Bud'ko and P. C. Canfield

*Ames Laboratory and Department of Physics & Astronomy,*

*Iowa State University, Ames, IA 50011*

1 February 2009

## Abstract

We discuss the irreversible magnetic properties of self-flux grown Ba(Fe$_{1-x}$Co$_x$)$_2$As$_2$ single crystals for a wide range of concentrations covering the whole phase diagram from the underdoped to the overdoped regime, x=0.038, 0.047, 0.058, 0.071, 0.074, 0.10, 0.106 and 0.118. Samples were characterized by a magneto-optical method and show excellent spatial uniformity of the superconducting state down to at least the micrometer scale. The in-plane properties are isotropic, as expected for the tetragonal symmetry, and the overall behavior closely follows classical Bean model of the critical state. The field-dependent magnetization exhibits second peak at a temperature and doping - dependent magnetic field, $H_p(T,x)$. The evolution of this fishtail feature with doping is discussed. In particular we find that $H_p$, measured at the same reduced temperature for different x, is a unique monotonic function of the superconducting transition temperature, $T_c(x)$, across all dopings. Magnetic relaxation is time-logarithmic and unusually fast. Similar to cuprates, there is an apparent crossover from collective elastic to plastic flux creep above $H_p$. At high fields, the field dependence of the relaxation rate becomes doping independent. We discuss our results in the framework of the weak collective pinning and show that vortex physics in iron-based pnictide crystals is much closer to high-$T_c$ cuprates than to conventional s-wave (including MgB$_2$) superconductors.

## Introduction

Superconductivity was recently discovered in polycrystalline LaFeAsO$_{1-x}$F$_x$ with a zero - field transition temperature of $T_c \approx 23$ K[1]. This breakthrough was followed by the realization of even higher $T_c$ values, as high as 55 K in RFeAsO$_x$F$_y$ ("1111" system with R=Nd,Sm,Pr) [2-5]. Soon after, superconductors based on the parent AFe$_2$As$_2$ system (abbreviated as "122", here A is an alkaline earth element, A = Ca,Sr, Ba) with $T_c$ up to 38 K were synthesized [6]. With such relatively high transition temperatures and an apparent chemical diversity, these iron-based pnictide superconductors have attracted much attention. In the 122 system, either the A or Fe sites can be doped to achieve superconductivity with holes or electrons as carriers.

There is an important difference between these two classes of pnictide superconductors. While single crystals of the oxygen-based 1111 system are very difficult to grow and they are still very small, large high-quality flux-grown single crystals of the oxygen-free 122 are available [7-9]. Studying single crystals is very important to determine the baseline properties of these materials unaffected by the extrinsic factors associated with polycrystalline materials, such as grain boundaries, morphological defects and uncertainty in the sample volume and internal structure. These factors often dominate the macroscopic electromagnetic response of the samples.

Thermodynamic, transport and electromagnetic properties of the 122 crystals have been studied in detail in many publications [7-22] (and references therein). Recently, we have focused on the systematic study of the London penetration depth, $\lambda(T)$, in of Ba(Fe$_{1-x}$Co$_x$)$_2$As$_2$ ("FeCo-122") with various cobalt doping levels, x [11, 12] (see also this special issue of Physica C). To our surprise, we have found an almost universal unconventional behavior of the low-temperature variation of $\lambda(T) \propto T^n$ with the exponent $n$ varying from 2 in the underdoped regime to about 2.5 in the overdoped regime. In addition, an indication of a sudden decrease in the superfluid density below optimal doping was found. Furthermore, in the FeCo-122 system, the underdoped samples also exhibit structural/magnetic transitions at temperatures above $T_c$ [8]. A detailed study of the vortex properties performed on nearly optimally doped (x=0.074) single crystals found a great deal of similarity between this material and clean high-$T_c$ cuprates. In particular, a fishtail feature (first reported for (Ba$_{1-x}$K$_x$)Fe$_2$As$_2$ , "BaK-122", system with x=0.4 [19]), very fast time-logarithmic magnetic relaxation and an apparent crossover from collective elastic to plastic creep [15]. With all these observations, the natural question is how do the vortex properties evolve when the doping level changes?

In this contribution, we first discuss characterization of the superconducting properties of Ba(Fe$_{1-x}$Co$_x$)$_2$As$_2$ single crystals with various x and show that all samples exhibit robust and spatially homogeneous superconductivity as revealed by 1) direct magneto-optical imaging of the Meissner screening; 2) visualization of trapped magnetic flux; 3) direct magnetization measurements of the superconducting transition. We then study the evolution of the fishtail feature when, in addition to the usual peak in the vicinity of $H=0$, a second peak appears in $M(H)$ at the magnetic field $H_p$. We find that $H_p(T/T_c = const)$ is a single – valued monotonic function of $T_c$. Furthermore, we find that magnetic relaxation is non-monotonic function of magnetic field, indicating a crossover to the plastic creep regime. At $H > H_p$, the logarithmic relaxation rate, $S = -|d\ln M/d\ln t|$, increases with applied magnetic field at a rate independent of doping level. Our observations, together with other measurements and reports [19, 22], point to unconventional irreversible vortex properties of iron-arsenides, at least as represented by the 122 family. They point to a close similarity with the cuprates and distinctly different from the behavior found in conventional s-wave superconductors, including two-gap MgB$_2$.

## Experimental

### Samples and characterization

Single crystals of Ba(Fe$_{1-x}$Co$_x$)$_2$As$_2$ with x=0.038, 0.047, 0.058, 0.071, 0.074, 0.10, 0.106 and 0.118 were grown out of self-flux (FeAs) [8]. The actual cobalt concentration was determined by wavelength dispersive x-ray spectroscopy in the electron probe microanalyzer of a JEOL JXA-8200 superprobe. The superconducting transition temperature, $T_c$, has been determined from the onset of a diamagnetic signal as well as from temperature when resistance became zero. It ranged from 8 to 24 to K. The quality of the samples was checked with magneto-optical imaging as described below in detail. Extensive thermodynamic and transport studies of the crystals from the same batches was reported elsewhere [8].

Conventional characterization was done using a commercial magnetometer (*Quantum Design* MPMS) and general purpose systems for specific heat and transport measurements (*Quantum Design* PPMS).

### Magneto-optical imaging

Magneto-optical imaging of the component of the magnetic induction perpendicular to the sample surface was conducted by utilizing the Faraday effect in bismuth-doped iron garnet ferrimagnetic films with in-plane magnetization [23]. Such a film is grown on the transparent substrate and then covered by a thin metallic layer to serve as a mirror. The whole structure is called an "in-plane magneto-optical indicator". When linearly polarized light passes through the indicator and reflects off of the mirror sputtered on its bottom, it picks up a double Faraday rotation proportional to the intensity of the magnetization along the light path – perpendicular to the indicator (and sample) surface. This component of magnetization, in turn, is proportional to the perpendicular component of the magnetic induction at a given location on the sample surface. Observed through the (almost) crossed (with respect to polarizer) analyzer, we recover a real-time 2D image where the intensity is proportional to the magnetic field on the sample surface [24]. To study superconductors, a flow-type liquid $^4$He cryostat with the sample in vacuum was used. The sample was positioned on top of a polished copper cold finger and an indicator was placed on top of the sample. In the experiment, the indicator is placed with the active side down in direct contact with the flat surface of the sample. It is the distance between the surface of the active layer and layer thickness itself that mostly determine the spatial resolution of the technique. Without special contrivances, we obtain spatial resolution of about 2 to 4 µm. The cryostat was positioned under the polarized-light reflection microscope and the color images could be recorded on video and high-resolution CCD cameras. Note that some images contain various "defects" – spots (sometimes bright) and streaks. These are optical artifacts due to dirt, grease, dust and scratches on the substrate or mirror and they do not affect the underlying image of magnetic field in any way.

## Results

### Sample characterization

Figure 1 (a) shows screening of a 1.5 kOe applied magnetic field at 5 K after cooling in zero field in a nearly optimally doped Ba(Fe$_{1-x}$Co$_x$)$_2$As$_2$ ($x = 0.074$) single crystal. Figure 1 (b) was obtained after the magnetic field was turned off and some flux was trapped at the sample perimeter. Figure 1 (c) shows an optical image of the sample and Figure 1 (d) shows trapped magnetic flux after the sample was cooled in a 1.5 kOe magnetic field to 5 K and the field was turned off. These observations clearly show robust and strong superconductivity of FeCo-122 superconductors with very uniform Meissner screening and vortex pinning, at least down to the micrometer scale set by the optical resolution of the magneto-optical imaging. The superconducting transitions of the six samples with x=0.038, 0.047, 0.058, 0.074, 0.10, and 0.118 are shown in Figure 2, where $M(T)$ measurements were performed upon warming after cooling in zero field and applying a 10 Oe field at 5 K. The superconducting transition temperature, $T_c$, shifts dramatically with doping, however unlike many previous reports, the transition does not broaden for all concentrations, probably reflecting the chemical homogeneity of the selected samples. (We note that we "screened" many samples before actual measurements and sometimes observed irregular behavior or broad transitions. Such samples were discarded.) The transition temperature versus measured cobalt concentration is summarized in Figure 3. Similar to other superconducting systems where superconductivity is induced by chemical doping, $T_c(x)$ has a rough dome shape with the maximum somewhere around x=0.065. Previous studies have shown that in FeCo-122 crystals, underdoped samples also exhibit structural/magnetic transitions from orthorhombic/AFM to the tetragonal/paramagnetic phase at temperatures above $T_c$ [8, 11]. This coexistence adds an interesting peculiarity stimulating studies over the whole doping phase diagram.

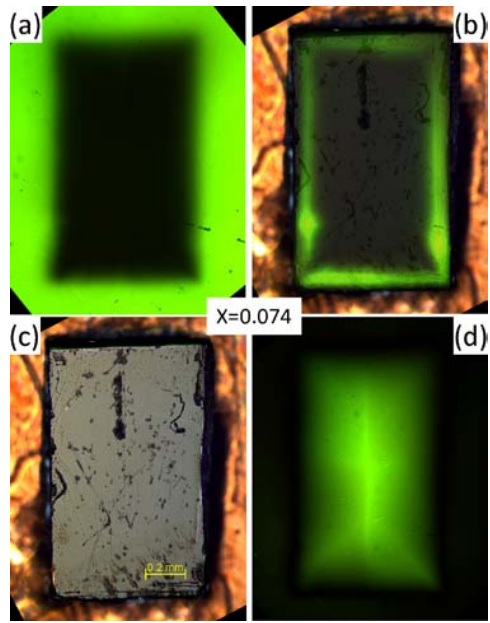

**Figure 1.** A nearly optimally doped sample with x=0.074 ($T_c = 22.8$ K). (a) Meissner screening of a 1.5 kOe applied magnetic field. (b) trapped flux after 1.5 kOe was turned off. (c) optical image of a sample (d) trapped flux after cooling in 1.5 kOe and turning it off.

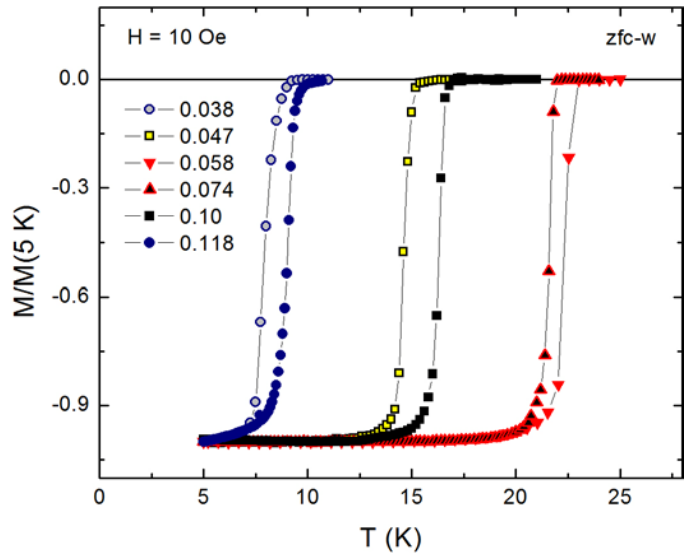

**Figure 2.** Superconducting transition temperatures for six different samples with doping level x shown in the legend.

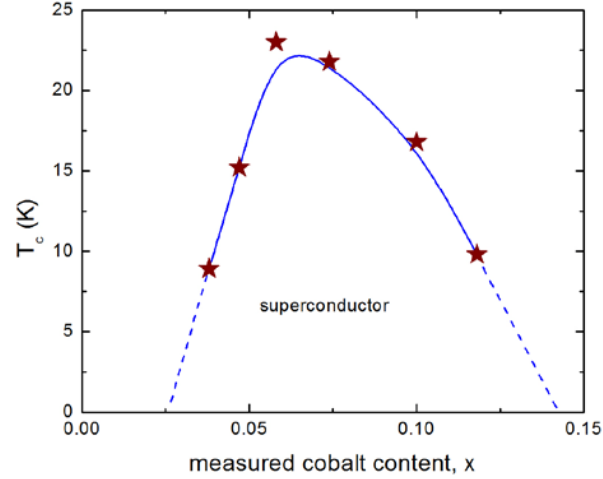

**Figure 3.** $T_c$ vs. x in Ba(Fe$_{1-x}$Co$_x$)$_2$As$_2$ single crystals.

Measurements of the penetration depth, $\lambda(T)$, at low temperatures have found non-exponential behavior, $\lambda(T) \propto T^n$, with the exponent $n$ changing from about 2 in the underdoped samples to almost 2.5 in the overdoped samples. At the same time, the superfluid density decreases abruptly in the underdoped regime. It is therefore important to study the vortex response for any possible "asymmetry" with respect to the doping level, x. To persue this goal, we first characterized the samples with non-optimal concentration for spatial uniformity and homogeneity of the superconducting state.

Figure 4 shows screening of an applied magnetic field (1 kOe at 5 K) in four samples with x=0.047, 0.071, 0.100 and 0.106. No defects or abnormal flux penetration were observed. The screening is very uniform across the doping indicating strong homogeneous shielding of fairly strong magnetic fields. Similarly, Figure 5 examines the structure of the magnetic flux trapped after cooling in a 1.5 kOe applied magnetic field from above $T_c$ to 5 K and turning the magnetic field off. The patterns are typical for type-II superconductors with homogeneous pinning. No apparent faults in the bulk or at the edges can be seen. Another important point to be made is the existence of in-plane anisotropy. It is not expected in a tetragonal system, but different growth defects and other morphological features may induce an anisotropy in pinning that will result in anisotropic trapping and screening of the magnetic flux. Figure 6 shows four stages of magnetic field penetration into a nearly optimally doped (x=0.074) single crystal. The lower right frame shows overlayed schematics of what is expected from a simple isotropic Bean critical state. A similar shape is clearly seen in the distribution of trapped flux, as shown in the lower right corner of Figure 5 (x=0.106). The expectation is very close to the observed pattern. (Note that the tooth-shaped overlay of darker and brighter areas seen in the lower left image of Figure 5 (x=0.100) is not due to trapped flux, but is an artifact due to birefringence in this particular magneto-optical indicator with in-plane domains that are positioned at certain angles with respect to the polarization plane.)

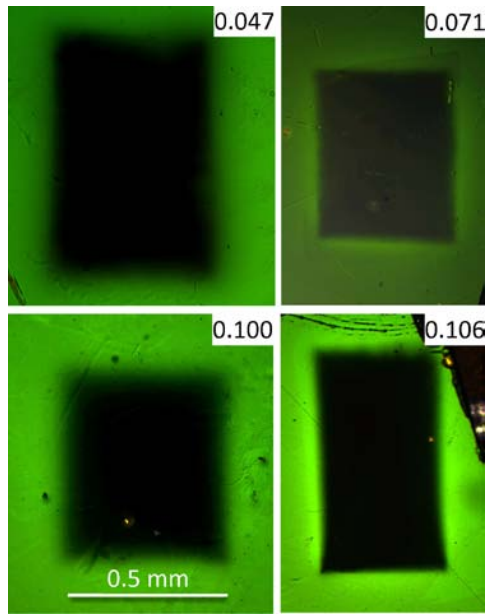

**Figure 4.** Meissner screening in single crystals with four indicated doping levels in magnetic fields of 1 kOe at $T = 5$ K.

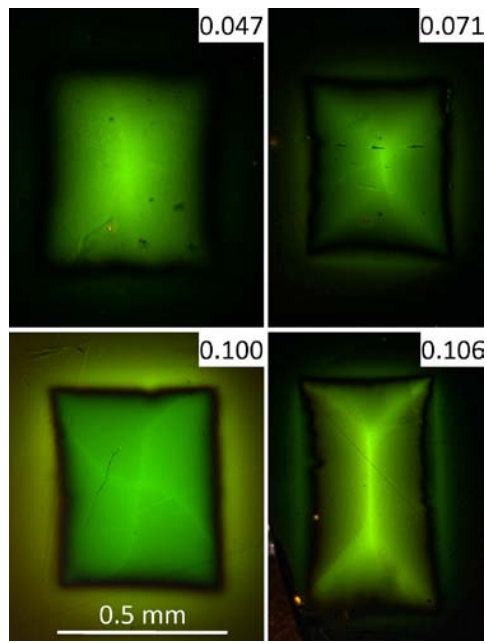

**Figure 5.** Magnetic flux trapped in samples after cooling in a 1.5 kOe magnetic field to 5 K and turning field off. Doping levels are indicated in right top corners.

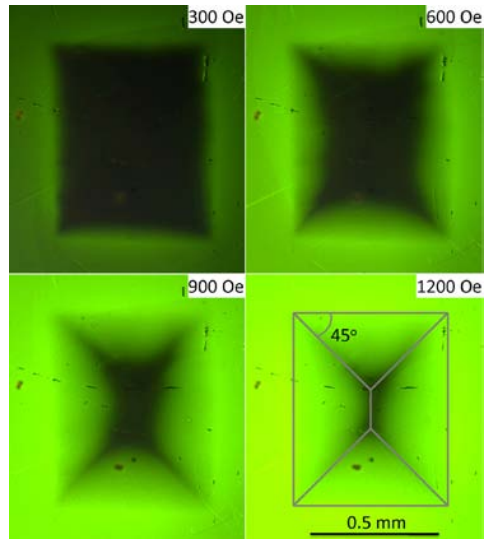

Figure 6.  Magnetic flux penetration at 20 K into a crystal with x=0.07. The last frame shows a schematic overlay of the expected "Bean oblique wedge" shape with isotropic in-plane current density.

We conclude by stating that magneto-optical imaging and magnetization measurements of the superconducting transition indicate high quality of our single crystals across a wide range of cobalt concentrations corresponding to a significant variation in the transition temperature on both underdoped and overdoped sides of the dome.

## Magnetization

All measurements reported in this paper were obtained with the applied magnetic field oriented along the crystallographic c-axis (usually along the shortest sample dimension). For discussion of anisotropic properties with other orientations of the magnetic field, see Refs.[15, 16]. Figure 7 shows magnetization loops measured at 5, 10 and 15 K in a Ba(Fe$_{1-x}$Co$_x$)$_2$As$_2$ single crystal with x=0.074, close to optimal doping (the crystal is shown in Figure 1 (c) with its c-axis perpendicular to the page). The same crystal was characterized in detail in Ref.[15]. Here we would like to point out the nonmonotonic feature called a fishtail and illustrate the (visual) definition of the characteristic magnetic field, $H_p(T)$, corresponding to the maximum in the magnetic moment.

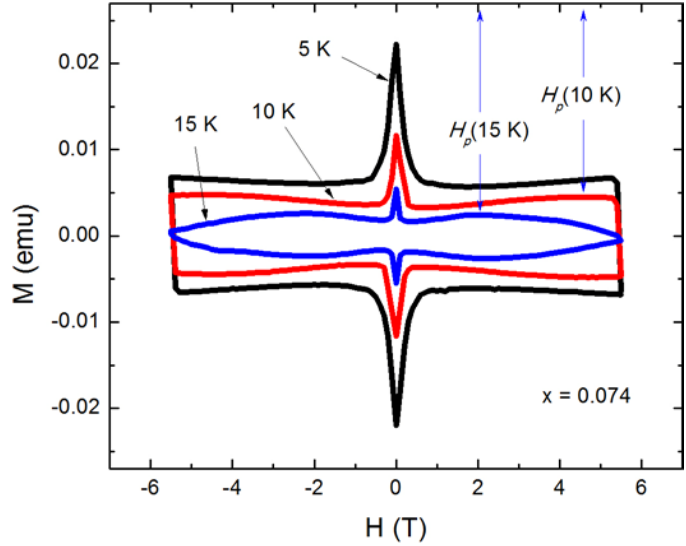

Figure 7. Magnetization loops measured in a Ba(Fe$_{1-x}$Co$_x$)$_2$As$_2$ single crystal shown in Figure 1 (c) for the magnetic field parallel to the c-axis (perpendicular to the page in Figure 1 (c)) at 5, 10 and 15 K. Also shown is the definition of the characteristic magnetic field $H_p$.

Similar to findings in the cuprates, $H_p(T)$ decreases with the increase of temperature and the non-monotonic behavior becomes more prominent. In order to compare magnetization loops between samples with different dopings, we have performed measurements at temperatures corresponding to the same reduced temperature of $t = T/T_c = 0.7$. The result is shown in Figure 8. While the lowest doping of x=0.38 ($T_c \approx 9\,\text{K}$) does not show fishtail magnetization, other concentrations do, but with a widely varying characteristic field, $H_p$. Note that x=0.058 ($T_c \approx 23.0\,\text{K}$) is closer to optimal doping than x=0.074 ($T_c \approx 21.8\,\text{K}$) and its $H_p$ is apparently larger.

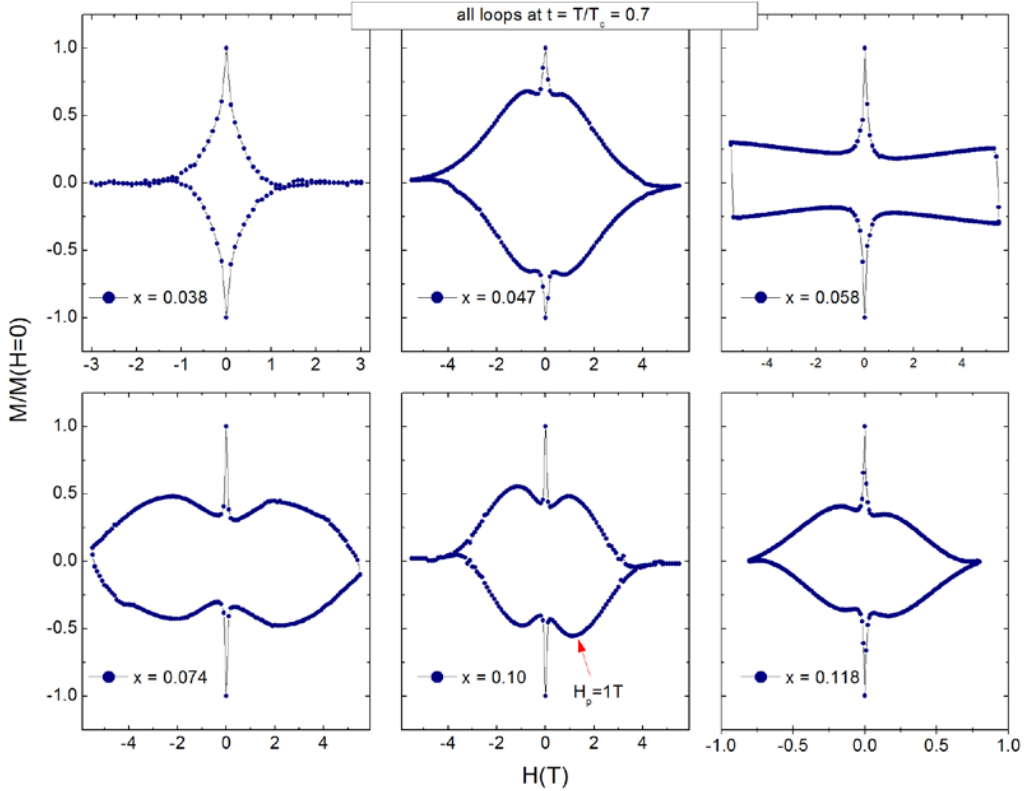

**Figure 8.** Magnetization loops measured in samples with indicated doping levels at the same reduced temperature $t = T/T_c = 0.7$.

To examine how sensitive the position of the peak is to the doping level x, we plot in Figure 9 the transition temperature as a function of $\ln(H_p)$ at a constant reduced temperature, $t = T/T_c = 0.7$. Despite the fact that $T_c(x)$ is a highly non-monotonic function (see Figure 2), the $T_c[\ln(H_p)]$ dependence is practically a straight line. This not only reflects an exponential dependence of $H_p$ on $T_c(x)$ at $t = const$, but more importantly, it shows that as far as pinning properties and the fishtail feature are concerned, both underdoped and overdoped samples behave similarly. This important conclusion will be reinforced when we discuss magnetic relaxation.

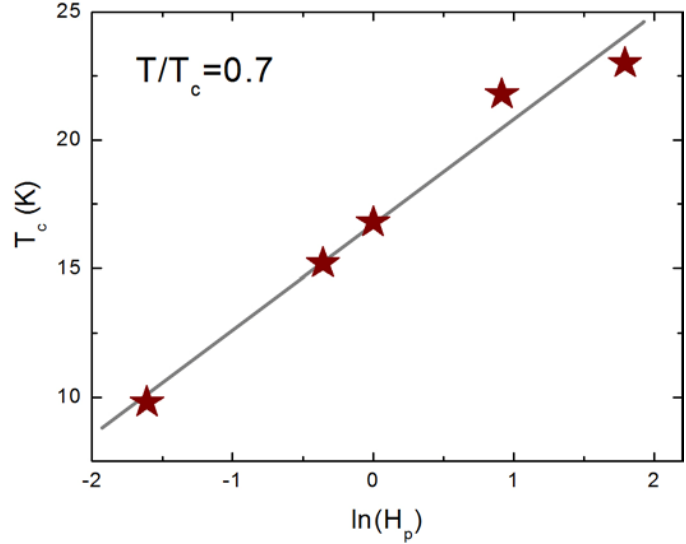

**Figure 9.** $T_c(H_p)$ presented on a log scale at $t = 0.7$ for samples shown in Figure 8.

## Magnetic relaxation

We now turn to a discussion of the flux dynamics. As reported earlier for nearly optimally doped FeCo-122, magnetic relaxation in this class of superconductors is almost time-logarithmic with a very large relaxation rate, $S = -|d\ln M/d\ln t|$ (see the inset in Figure 10 for raw data), comparable to or larger than in the cuprates [15]. This fact alone indicates unconventional vortex behavior with fluctuations playing an important role. This may also explain a substantial variation of the "critical" current density reported in the literature. The actual value of the measured supercurrent density is determined by the experimental time-window and different methods will find different values. Direct comparison of transport and magnetic current densities in FeCo-122 has been reported elsewhere [15, 16].

An obvious region of interest is where the fishtail effect is clearly seen. Figure 10 shows a set of measurements performed in different magnetic fields at 15 K in Ba(Fe$_{1-x}$Co$_x$)$_2$As$_2$ crystal with x=0.074, close to optimal doping. For each run, the magnetic field was first ramped to negative 5 T and then increased to a target value at which time the measurement started. With magnetometers, such as Quantum Design MPMS, there is always some ambiguity of the initial time of the measurement (and each measurement takes a few seconds). We therefore have measured the magnetic relaxation of the same sample under similar conditions in a magneto-optical setup where the initial time of 20 msec was well-controlled. In this experiment, the waiting time was long enough to overlap with the MPMS data. By matching the relaxation curves, we found that the initial time for the MPMS is about 50 sec, at least for the measurement protocol described above.

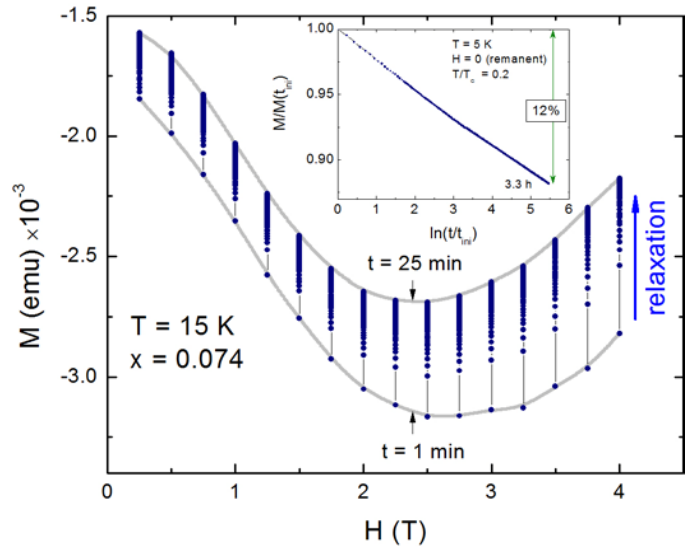

Figure 10. Relaxation of a magnetic moment measured over 25 minutes at 15 K in a Ba(Fe$_{1-x}$Co$_x$)$_2$As$_2$ single crystal with x=0.074 at different values of applied magnetic fields. Inset: 3.3-hour relaxation showing almost perfect time-logarithmic behavior of $M(time)$.

As can be seen in Figure 10, the magnetic relaxation is field dependent and causes $H_p$ to shift to lower values with time. In order to examine the field dependence we have evaluated the logarithmic relaxation rate at times much longer than the initial time, so that $\ln M$ plotted vs. $\ln t$ is linear (usually about 5 minutes after the magnetic field was stabilized). The result is shown in Figure 11. At first, the relaxation rate decreases as expected from the collective pinning model where flux creep proceeds via reversible elastic deformations and the vortex bundle growth leads to an increase of the effective barrier for magnetic relaxation with increasing field [25]. However, in the vicinity of $H_p$ and above, the magnetic relaxation apparently accelerates with increasing magnetic field. Here, the situation is opposite to the collective creep case and it was explained by the crossover from the collective (elastic) flux creep to plastic creep where an elementary perturbation of vortex matter is due to motion of the dislocations in the vortex lattice [26-28].

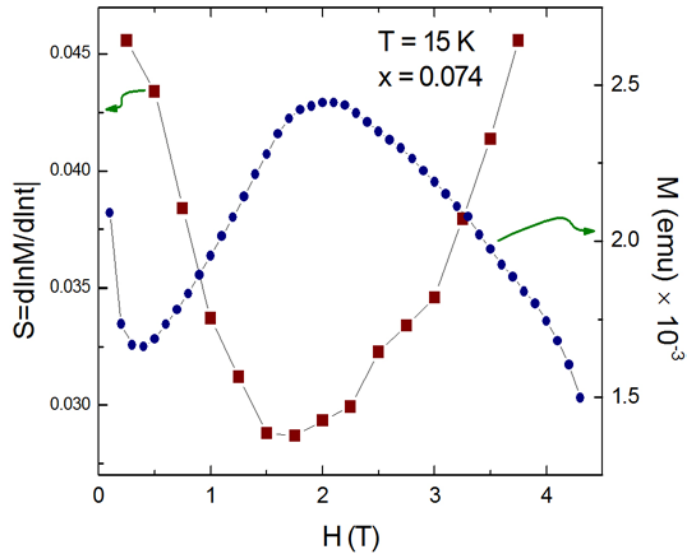

**Figure 11. Logarithmic relaxation rate (left axis) shown along with part of the $M(H)$ loop indicating correlated non-monotonic behavior.**

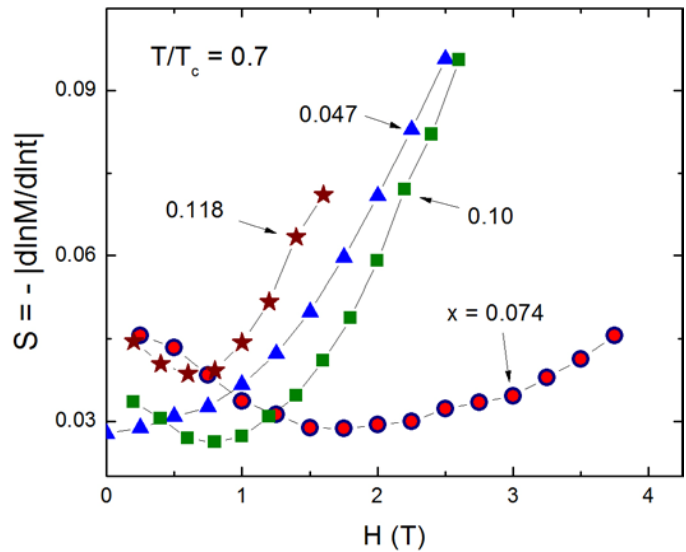

**Figure 12. Magnetic field dependence of the logarithmic relaxation rate for four different concentrations.**

We finally compare the field-dependent relaxation rates for samples of different dopings measured at the same reduced temperature, $t = 0.7$. In all cases, $S$ increases with increasing magnetic field above $H_p$, however it is interesting to note that the slope, $dS/dH$, reaches a universal value, both for underdoped and overdoped samples and we believe the slope of the optimally doped sample will reach this value at higher fields.

This aspect of vortex dynamics is apparently independent of doping and is probably related to the elementary mechanism of plastic relaxation.

## Conclusions

It was shown that Ba(Fe$_{1-x}$Co$_x$)$_2$As$_2$ single crystals exhibit robust and spatially homogeneous superconductivity for all doping levels studied (covering $T_c$ from 24 K down to 8 K on both underdoped and overdoped sides of the superconducting dome). Analysis of irreversible static and dynamic magnetic properties, in particular non-monotonic fishtail feature and fast field-dependent magnetic relaxation, point to a close similarity with the high-$T_c$ cuprates. Moreover, the vortex behavior of iron arsenides is as anisotropic as that of the cuprates [15, 16]. On the other hand, irreversible magnetic properties of the pnictides are distinctly different from conventional s-wave superconductors and two-gap MgB$_2$. This similarity to the cuprates is striking, considering weak electronic anisotropy of iron arsenides [16]. We speculate that several major factors could be responsible for these unusual properties. Both classes are layered, which can significantly affect the anisotropy of vortex pinning. In addition, superconductivity in both systems is in proximity to magnetism. Taking into account highly anisotropic magnetic interactions, the anisotropy of irreversible properties may not be that surprising.

## Acknowledgements

We thank J. R. Clem, V. G. Kogan and A. E. Koshelev for useful discussions and comments. Work at the Ames Laboratory was supported by the Department of Energy-Basic Energy Sciences under Contract No. DE-AC02-07CH11358. R. P. acknowledges support from Alfred P. Sloan Foundation. M.A.T. acknowledges continuing cross-appointment with Institute of Surface Chemistry, NAS Ukraine.